\documentclass{Interspeech}

\interspeechcameraready

\usepackage{multicol}
\usepackage{multirow, makecell}

\title{Approaching Dialogue State Tracking via Aligning Speech Encoders and LLMs}

\author[affiliation={1}]{Šimon}{Sedláček}
\author[affiliation={1}]{Bolaji}{Yusuf}
\author[affiliation={1}]{Ján}{Švec}
\author[affiliation={1,2}]{Pradyoth}{Hegde}
\author[affiliation={1}]{Santosh}{Kesiraju}
\author[affiliation={1}]{Oldřich}{Plchot}
\author[affiliation={1}]{Jan}{Černocký}

\affiliation{Speech@FIT}{Brno University of Technology}{Czechia}
\affiliation{}{Indian Institute of Information Technology Dharwad}{India}
\email{\{isedlacek,iyusuf,isvecjan,kesiraju,iplchot,cernocky\}@fit.vut.cz, pradyothhegde@gmail.com}

\usepackage{comment}
\usepackage{tikz}
\usetikzlibrary{shapes.geometric, arrows}
\usetikzlibrary{positioning, calc}
\usetikzlibrary{bayesnet}
\usetikzlibrary{shadows,shadings,shapes.symbols}

\makeatletter
\def\bstctlcite{\@ifnextchar[{\@bstctlcite}{\@bstctlcite[@auxout]}}
\def\@bstctlcite[#1]#2{\@bsphack
  \@for\@citeb:=#2\do{%
    \edef\@citeb{\expandafter\@firstofone\@citeb}%
    \if@filesw\immediate\write\csname #1\endcsname{\string\citation{\@citeb}}\fi}%
  \@esphack}
\makeatother

\begin{document}
\newcommand\todo[1]{\textcolor{red}{[todo: {#1}]}}
\bstctlcite{IEEEexample:BSTcontrol}

\maketitle

\begin{abstract}
In this work, we approach spoken Dialogue State Tracking (DST) by bridging the representation spaces of speech encoders and LLMs via a~small connector module, with a~focus on fully open-sourced and open-data components (WavLM-large, OLMo). We focus on ablating different aspects of such systems including full/LoRA adapter fine-tuning, the effect of agent turns in the dialogue history, as well as fuzzy matching-based output post-processing, which greatly improves performance of our systems on named entities in the dialogue slot values. We conduct our experiments on the SpokenWOZ dataset, and additionally utilize the Speech-Aware MultiWOZ dataset to augment our training data. Ultimately, our best-performing WavLM + connector + OLMo-1B aligned models achieve state of the art on the SpokenWOZ test set (34.66\% JGA), and our system with Gemma-2-9B-instruct further surpasses this result, reaching 42.17\% JGA on SpokenWOZ test.
    
\end{abstract}

\newcommand\bolaji[1]{\textcolor{blue}{[Bolaji: {#1}]}}

\section{Introduction}



Task-oriented dialogues (ToD) are multiturn conversations between a user and an agent, where the former has a specific goal (e.g., booking a restaurant for 5 people on Friday night) that is achieved with the help of the agent.
A key component of automated ToD systems is dialogue state tracking (DST), the task of tracking the user's intent (e.g. ``book-restaurant") and identifying the slots (e.g. ``restaurant-people": ``5", ``day": ``Friday").
Recently, aided by the DSTC-11 challenge~\cite{soltau-etal-2023-dstc} and the development of realistic SpokenWoZ~\cite{spokenwoz_2023} dataset, the scope of DST research is slowly advancing from being exclusively text-based to speech domain.

A typical approach to DST from spoken conversations is via a cascade of several systems: automatic speech recognition (ASR) $\rightarrow$ error correction module $\rightarrow$ text-based DST \cite{soltau-etal-2023-dstc,jiang-etal-2023-speech}.
Although end-to-end (E2E) DST training offers an attractive alternative due to the simplicity of training and the potential to avoid cascading errors, E2E DST models are difficult to train due to the scarcity of data compared to other speech processing tasks.
The paradigm of interconnecting pre-trained speech encoders with large language models (LLMs) offers a solution to the issue and has shown performance that is competitive with cascade systems~\cite{ReSLM:2024}.
However, most prior work relies on fully or partly closed models for which either the model weights or training data are not openly available.

Modality matching is one of the key challenges in aligning pre-trained speech and language models. Various approaches such as text-representation up-sampling \cite{chen22r_interspeech}, speech-representation sub-sampling and fixed-length representations based on Q-formers have been studied in the past \cite{Connector:Wenyi:2024, simon:2024:align}. The application of such speech language models was not only studied on traditional tasks such as ASR, speech translation and text-to-speech \cite{tts4,LLM_E2E_ASR:2024}, but also on language understanding. question-answering tasks and dialogue state tracking \cite{SLM:2023,tang2024salmonn,dong24_interspeech}. Retaining the ability of pre-trained LLMs even after the alignment is one of the main challenges \cite{ReSLM:2024}, as simple connector based approaches \cite{ma2024:kiss} do not generalize well beyond the domain of the training data \cite{kumar:2025:slam_asr}.
 Our work complements the prior works, as we study the task of dialogue state tracking from spoken conversations with fully open speech and language models.
 
In this paper, we propose an end-to-end DST model based on fully open components.
We connect a WavLM speech~\cite{wavlm_2022} encoder with an OLMo~\cite{groeneveld-etal-2024-olmo} LM through a small trainable connector module, which we first train to align the representation spaces of the pre-trained modes and then fine-tune the resulting model directly for DST.
We conducted experiments on the SpokenWoZ~\cite{spokenwoz_2023} and Speech-Aware MultiWoZ~\cite{soltau-etal-2023-dstc} datasets, showing that our proposed system with the fully open OLMo-1B model already yields state-of-the-art performance on SpokenWoZ and that we can obtain further significant improvements when we use the Gemma-2-9B-Instruct model.

\section{Method overview}
Our method treats DST as the problem of mapping from user speech and user-agent dialogue history to a~dictionary containing the correct transcription of the user's speech, and the inferred dialogue states (domains, slots). Therefore, we adopt a~model which we train to take speech as input and directly output a~JSON string representing this dictionary at each turn of the dialogue.

The model is composed of three parts: a~pretrained speech encoder, a~pretrained LLM and a~small connector module that joins the speech encoder and the LLM. The encoder computes a~representation of the speech input at the current turn of the conversation. Then, the speech representations are downsampled to better match the granularity of the LLM text input, and connector module maps them into the text embedding space of the LM, where they act as soft prompts.
Finally, the LM auto-regressively generates the tokens that comprise the JSON string. The LM input is prefixed with the connector output to provide speech conditioning and by the dialogue history in the form of the transcription of previous user inputs and agent responses.

\subsection{Training}
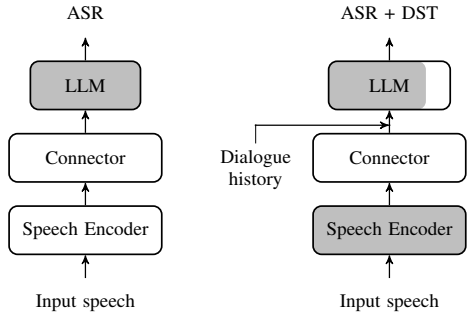
\begin{figure}[!t]
    \centering
    \scalebox{0.8}{
        \begin{tikzpicture}
	[	
	rect/.style={minimum width=22mm,minimum height=8mm,text width=23mm,
		align=center, rectangle,draw,thick, rounded corners, fill=gray!50},
	rect2/.style={minimum width=22mm,minimum height=8mm,text width=23mm,
		align=center, rectangle,draw,thick, rounded corners, fill=white},
        rect3/.style={minimum width=20mm,minimum height=8mm,text width=16mm,
		align=center, rectangle,draw,thick, rounded corners, fill=white},
        var/.style={minimum width=20mm,minimum height=8mm,text width=25mm,       
		align=center,rectangle,align=center},
        hidden/.style={circle,scale=0.05,minimum size=1pt,draw},
        arr/.style={->,>=stealth',semithick},
        fill fraction/.style n args={2}{path picture={
            \fill[#1] (path picture bounding box.south west) rectangle
            ($(path picture bounding box.north west)!#2!(path picture bounding box.north east)$);}}
        ]
        \node (inp) [var] at (0, -2.2) {Input speech};
	\node (enc) [rect2]  at (0, -1)  {Speech Encoder};
	\node (con) [rect2]  at (0, 0.2)  {Connector};
        \node (llm) [rect, minimum width=16mm,text width=16mm]  at (0, 1.4)  {LLM}; 
        \node (out) [var]  at (0, 2.6) {ASR}; 
        \node (inp2) [var] at (5, -2.2) {Input speech};        
	\node (enc2) [rect]  at (5, -1)  {Speech Encoder};
	\node (con2) [rect2]  at (5, 0.2)  {Connector};
        \node (inpd) [var] at (2.8, 0.0) {Dialogue \\ history};
        \node (h1) [hidden] at (2.8, 0.74)   {};
        \node (h2) [hidden] at (5, 0.74)   {};
        \node (llm2) [rect3, fill fraction={gray!50}{0.8}]  at (5, 1.4)  {LLM}; 
        \node (out2) [var]  at (5, 2.6) {ASR + DST}; 
        \draw [arr] (inp) edge (enc);
        \draw [arr] (enc) edge (con);
        \draw [arr] (con) edge (llm);
        \draw [arr] (llm) edge (out);
        \draw [arr] (inp2) edge (enc2);
        \draw [arr] (enc2) edge (con2);
        \draw [arr] (con2) edge (llm2);
        \draw [arr] (llm2) edge (out2);
        \draw (inpd) edge (h1);
        \draw [arr] (h1) edge (h2);
    \end{tikzpicture}
    }
    \caption{(Left) Stage-1 pre-training for ASR. (Right) Stage-2 training for joint ASR and dialogue state tracking (DST). Shaded modules are frozen during training. Partly shaded implies that the module is frozen, but additional trainable parameters are added via LoRA.}
    \label{fig:arch}
    \vspace{-2mm}
\end{figure}

Starting from pretrained encoder, we train the model in the two stages illustrated in Figure~\ref{fig:arch}: an ASR pretraining stage and a~joint ASR-DST finetuning.
\subsubsection{ASR pre-training}
Since the speech encoder and LM are pretrained independently of each other, we first bridge their representation spaces before fine-tuning for DST.
In this first stage, we freeze the LM and finetune the encoder and connector for ASR, conditioning the LLM solely using the connector outputs\footnote{\scriptsize ASR prompt format: `\%speech\_embeds\% \{``transcription": \%labels\%\}'}.

Thus, we are able to leverage large scale ASR datasets (which are orders of magnitude larger than typical DST datasets) to obtain robust alignment of the encoder and the LM.
Moreover, ASR training is necessary since the dialogue history for the final DST model requires transcribing the user input.

\subsubsection{Joint ASR-DST finetuning}
In the second training stage, we keep the speech encoder frozen, and introduce LoRA~\cite{hu2022_lora} adapters into the LM, which we finetune along with the connector for the target DST task.

For each turn in a~conversation, we append the ground-truth dialogue history to the connector output speech embeddings, and train the model to minimize the negative log-likelihood of the JSON string encoding the dialogue state for the current turn -- the transcription, domains, and slots.

Note that it is imperative that we still train jointly for ASR and DST since the model will not have access to the full ground-truth dialogue history at inference time, and will have to rely on its own ASR capabilities to get the user side of the dialogue.

\subsection{Inference and post-processing}
Inference proceeds in a~turn-by-turn fashion.
For each turn, the JSON string outputted by the model is converted into the corresponding dictionary whose elements are read as the dialogue state and the dictionary field corresponding to the user ASR transcription is appended to the previous dialogue history for use in subsequent turns of the dialogue.



To obtain the final dialogue slots, we employ a~fuzzy matching\footnote{\texttt{\url{pypi.org/project/fuzzywuzzy/}}} scheme, which is typically used in the DST community~\cite{jiang-etal-2023-speech}. This scheme maps the given slot value to the closest one in the database.

\section{Experiments}
\subsection{Datasets}
We conduct our experiments using two speech-grounded task-oriented dialogue datasets: SpokenWoZ~\cite{spokenwoz_2023} (SWOZ) and Speech-Aware MultiWoZ~\cite{soltau-etal-2023-dstc} (MWOZ), built on top of the original text-based MultiWoZ 2.1~\cite{budzianowski-etal-2018-multiwoz,eric-etal-2020-multiwoz}. We remove nine originally corrupted conversations from the SpokenWoZ test set\footnote{\url{github.com/AlibabaResearch/DAMO-ConvAI/issues/87}}. For SpokenWOZ, we also generate ASR transcripts using Whisper-large-v3~\cite{radford_robust_2023} both to serve as better reference text
for training the end-to-end system, and as input to our cascaded systems.
For evaluation, we adopt the standard MWOZ evaluation script~\cite{nekvinda-dusek-2021-shades}\footnote{\texttt{\url{github.com/Tomiinek/MultiWOZ_Evaluation}}} and report Joint Goal Accuracy (JGA) as well as Slot Error Rate (SER) for all test sets.

For the speech encoder/connector ASR pre-training stage, we use the Fisher-Switchboard (2000h)~\cite{godfrey_switchboard_1992} Librispeech (1000h)~\cite{panayotov_librispeech_2015} and How2 (300h)~\cite{how2:2018} datasets.

\subsection{Models and training configuration}

We use WavLM-large\footnote{\url{huggingface.co/microsoft/wavlm-large}}~\cite{wavlm_2022} pre-trained on 96k hours of speech data (Libri-Light, VoxPopuli and GigaSpeech), as the speech encoder in our experiments.
We primarily use OLMo-1B~\cite{groeneveld-etal-2024-olmo} as the pretrained language model.
We choose this model to address test set contamination concerns, as their training data are openly documented.
Additionally, we conduct some experiments using Gemma-2-9B-Instruct\footnote{\url{huggingface.co/google/gemma-2-9b-it}}~\cite{gemma2024, gemma2} as the base LM in order to facilitate comparison with some prior work.

The connector module is a~2-layer transformer encoder with 16 attention heads, hidden size 1024 and feedforward intermediate dimension of 4096.
This transformer is preceded by a~subsampling layer, which stacks 6 neighbouring WavLM embeddings and projects them into the hidden size of the connector, resulting in 6x downsampling.

In the ASR pretraining stage, we train with batch size of 64, learning rate of 1e-4 and 2000 warmup steps and train until the cross-entropy on the ASR dev sets stops improving.

In the joint ASR-DST training stage, we set the rank of the LoRA layers to r=16, and train with batch size of 128, learning rate of 5e-5 and 500 warmup steps on both SWOZ and MWOZ until the cross-entropy on the combined dev set stops improving.
Then, optionally, we perform one additional epoch on fine-tuning separately for each dataset, with a~batch size of 256 on the target dataset (referred to subsequently as FT-sw for SWOZ and FT-mw for MWOZ).

\subsection{Baselines and Cascaded Systems}
We establish a~number of baseline cascaded Whisper/OLMo-1B DST systems, experimenting with full fine-tuning of the foundation LLMs, LoRA adapters, effects of providing agent-side dialogue history, and post-processing.

When training, we compute loss over the whole dialogue prompt schema (even the history), when decoding, only the domain and slot predictions are generated. Whisper-large-v3 is used as the ASR frontend instead of the original SpokenWoZ transcripts. The best-performing cascaded system on SpokenWoZ test is shown in row three of Table \ref{tab:best_models}.

We also compare our results to the best system from~\cite{spokenwoz_2023} and the relevant corresponding Gemma-2-9B-Instruct system from~\cite{richardson:2024:schema_aug} (first two lines of Table \ref{tab:best_models}). Further analysis of our cascaded systems is available in Section \ref{sec:cascaded_analysis}.


\begin{table}[!tp]\centering
\caption{SpokenWoZ JGA comparison of our best-performing and cascaded and E2E models with two reference systems (1, 2) from prior works.}\label{tab:best_models}
\footnotesize
\begin{tabular}{lcc}\toprule
\multirow{2}{*}{\textbf{Model}} &\multicolumn{2}{c}{\textbf{SWOZ test}} \\\cmidrule{2-3}
&JGA &SER \\\midrule
(1) Gemma-2-9B-Instruct (cascaded)~\cite{richardson:2024:schema_aug} &25.40 &- \\
(2) SPACE+WavLM\_{align}~\cite{spokenwoz_2023} &25.65 &- \\
Whisper\textrightarrow OLMo-1B, full FT, SW+MW (cascade) &30.74 &31.11 \\
WavLM + conn. + OLMo-1B (A11 from Table \ref{tab:aligned}) &34.66 &26.80 \\
WavLM + conn. + Gemma-2-9B-Instruct (Table \ref{tab:gemma}) &\textbf{42.17} &\textbf{20.41} \\
\bottomrule
\end{tabular}
\end{table}

\subsection{Aligned systems}
\label{sec:aligned_experiments}
When training the aligned speech encoder + connector + LLMs systems, we explore different training and inference configurations and ablating their effects. This includes the datasets used for training, connector ASR pre-training, adding LoRA to the LLM, using just the user turns in the dialogue history (for user-agent turns we insert speaker tags `\texttt{USER:}'/`\texttt{AGENT:}'). 
The DST prompt format has the following structure for these experiments: \emph{\%speech\_embeds\% \{``dialogue\_history": \%context\%, ``current\_turn": \%asr\_hyp\%, ``domains": [\%domains\%], ``slots": \{\%slots\%\}\}}. For inference, the model completes the JSON starting with the ASR hypothesis, conditioned on the speech embeddings and the dialogue history.

We train the aligned systems with the primary goal of achieving best possible performance on SWOZ, however, we also evaluate on the MWOZ dev set to get a~more complete picture of the generalization of the model. Table \ref{tab:best_models} shows the best OLMo-1B aligned DST system we obtained (A11 from Table \ref{tab:aligned}). Further analysis and description of our resulting aligned systems can be found in section \ref{sec:aligned_analysis}.

\section{Results}
\label{sec:results}

\subsection{Cascaded system analysis}
\label{sec:cascaded_analysis}

Our experiments focus on ablating three main training setup factors: the training data used, the type of LLM fine-tuning, and the usage of both user and agent turns in the dialogue history. a~full overview of our cascaded systems is provided in Table \ref{tab:cascaded}. We additionally include two text-only baseline systems from DSTC-11~\cite{soltau-etal-2023-dstc} as reference baselines for speech-aware MWOZ.

\begin{table}[!tp]\centering
\caption{Cascaded Whisper + OLMo-1B baseline systems with two MWOZ dev reference systems from~\cite{soltau-etal-2023-dstc}}\label{tab:cascaded}
\scriptsize
\begin{tabular}{lcccc|cc}\toprule
\multicolumn{3}{c}{\textbf{Cascaded system configuraiton}} &\multicolumn{2}{c}{\textbf{SWOZ test}} &\multicolumn{2}{c}{\textbf{MWOZ dev}} \\\cmidrule{1-7}
Training data &FT &UA &JGA &SER &JGA &SER \\\midrule
SW-orig. &full &- &26.07 &38.71 &12.87 &49.52 \\
SW &full &- &28.81 &34.26 &12.49 &52.73 \\
MW &full &- &16.17 &58.15 &27.72 &30.34 \\
SW, MW &full &- &27.74 &34.51 &19.04 &40.89 \\
SW, MW &lora64 &- &26.66 &36.45 &19.76 &38.11 \\
\midrule
SW &full &yes &29.91 &31.93 &12.54 &57.69 \\
MW &full &yes &16.34 &58.84 &29.48 &27.39 \\
SW, MW &full &yes &\textbf{30.74} &\textbf{31.11} &23.37 &34.57 \\
\midrule
\multicolumn{3}{l}{ASR-DSTC large (770M params.)} & & &25.2 & \\
\multicolumn{3}{l}{ASR-DSTC xxl (11B params.)} & & &43.1 & \\
\bottomrule
\end{tabular}
\end{table}

First, we observe that fine-tuning on the original SpokenWoZ transcripts yields worse results than when using Whisper transcripts, which are of higher quality. Second, we find it beneficial to combine both SpokenWoZ and SA-MultiWOZ for training.

While full LLM fine-tuning yields the best results, we also experiment with LoRA adapters. We find that our cascaded systems achieve best results with bigger LoRA ranks, and we show the best cascade LoRA-64 system in Table \ref{tab:cascaded}. When scaling up the foundation LLMs or training E2E spoken DST systems, full fine-tuning becomes inconvenient. Therefore we treat the full FT cascades as the toplines for the E2E LoRA systems.
Lastly, while using only the user turns in the dialogue history yields good results, adding also the agent turns improves JGA on both datasets, resulting in our best cascaded SWOZ result of 30.74\% JGA without post-processing.

\subsection{Aligned system analysis}
\label{sec:aligned_analysis}

\begin{table*}[t]\centering
\caption{Results of our aligned WavLM + connector + OLMo-1B systems. UA denotes the usage of both user and agent turns in the dialogue history, FT-sw/FT-mw denotes additional final fine-tuning on the given dataset.}\label{tab:aligned}
\begin{tabular}{cclcccc}\toprule
\multirow{2}{*}{\textbf{\#}} &\multirowcell{2}{\textbf{Training} \\ \textbf{data}} &\multirow{2}{*}{\textbf{System configuration}} &\multicolumn{2}{c}{\textbf{SpokenWoZ test}} &\multicolumn{2}{c}{\textbf{MultiWOZ dev}} \\\cmidrule{4-7}
& & &JGA[\%] &SER[\%] &JGA[\%] &SER[\%] \\\midrule
A1 &\multirow{6}{*}{SW} &lora16 (no ASR init) &22.43 &44.46 &- &- \\
A2 & &connector-only &20.02 &54.55 &- &- \\
A3 & &lora16 &27.89 &33.23 &- &- \\
A4 & &lora16 + UA &29.97 &31.62 &- &- \\
A5 & &lora16 + UA + FUZZY &\textbf{32.08} &\textbf{29.50}&- &- \\
\midrule
A6 &\multirow{8}{*}{SW+MW} &lora16 (no ASR init) &23.32 &42.33 &13.32 &52.08 \\
A7 & &connector-only &17.58 &61.41 &10.75 &66.64 \\
A8 & &lora16 &27.27 &35.49 &16.69 &46.12 \\
A9 & &lora16 + UA &31.04 &29.82 &16.63 &44.57 \\
A10 & &lora16 + UA + FT-sw &31.91 &28.98 &- &- \\
A11 & &lora16 + UA + FT-sw + FUZZY &\textbf{34.66} &\textbf{26.80} &- &- \\
A12 & &lora16 + UA + FT-mw &- &- &17.49 &43.99 \\
\midrule
A13 &\multirow{5}{*}{MW} &connector-only &- &- &11.32 &60.53 \\
A14 & &lora16 &- &- &15.34 &44.96 \\
A15 & &lora16 + UA &- &- &18.20 &37.86 \\
A16 & &lora16 + UA + FUZZY &- &- &21.62 &34.29 \\
\bottomrule
\end{tabular}
\end{table*}

An overview of all our WavLM-large + connector + OLMo-1B aligned DST systems is shown in Table \ref{tab:aligned}.

We observe that initializing the connector and speech encoder randomly and fine-tuning with LoRA (A1, A6) leads to much better results, compared to only fine-tuning the connector (A2, A7, A13). However, the two step pre-training brings additional improvements when used jointly with LoRA. Also, using both the user and agent turns in the dialogue history brings consistent improvements in JGA (e.g. A9-12).

While training only on SpokenWoZ yields reasonable results, we observe a~strong tendency for overfitting on such a~small dataset. We therefore find it beneficial to add SA-MultiWOZ to the training set as apart from the \emph{profile} domain, MWOZ has the same schema as SWOZ. However, the training speech data is TTS, which is why we primarily regard this dataset as an augmentation of the DST part of the training. We obtain our best aligned OLMo-1B system by training on both datasets, and then fine-tuning on SpokenWoZ for one more epoch with a~batch size of 192, achieving 34.66\% JGA after post-processing (A11).

The ASR pre-training we perform focuses primarily on natural and conversational speech, and we find that our aligned models do not handle the MWOZ dev/test sets nearly as well. However, when ablating utilizing the MWOZ dev ground truth user history in the context of our model, we obtain a~significant improvement in JGA (18.2 to 27.1), indicating that the utilization of a~speech encoder better suited for the \emph{human} MultiWOZ data domain would help close the performance gap between the two datasets. Similarly, when using the Whisper context for inference on SpokenWoZ instead of the previously decoded user utterances , we obtain a~further increase in JGA from 31.91 (A10) to 32.89.

Lastly, we show the effect of LLM output post-processing using fuzzy matching of the slot values, which, on average, yields a~further steady improvement of ~3\% JGA absolute.

\subsection{Scaling the foundation models}
\label{sec:aligned_analysis}

Our goal with using the WavLM and OLMo models was predominantly transparency in terms of potential test data contamination. However, inspired by other works~\cite{richardson:2024:schema_aug} we perform additional experiments with the larger Gemma-2-9B-Instruct.
With the same WavLM + connector configuration and training hyperparameters as for our other models, we first run the ASR pre-training (this time with the encoder frozen as we find that unfreezing it does not yield significant improvements for larger LLMs), then we fine-tune the connector and respective LLM with LoRA r=8 for DST on both SWOZ and MWOZ.
The results are shown in \ref{tab:gemma}. 
Note that we also experimented with the OLMo-7B model but we omit it from these results as it did not outperform its A10/A11 OLMo-1B counterpart from Table \ref{tab:aligned}.

\begin{table}[!htp]\centering
\caption{WavLM-large + connector + Gemma-2-9B-Instruct aligned model results on the SWOZ tes and MWOZ dev/test sets.}\label{tab:gemma}
\begin{tabular}{lcc}\toprule
&JGA[\%] &SER[\%] \\\midrule
\multicolumn{3}{l}{\textbf{SWOZ test (FT-sw)}} \\
Gemma-2-9B-Instruct & 38.76 & 22.66 \\
Gemma-2-9B-Instruct + FUZZY & \textbf{42.17} & \textbf{20.41} \\
\midrule
\multicolumn{3}{l}{\textbf{MWOZ test human-verb. (FT-mw)}} \\
Gemma-2-9B-Instruct & 21.39 & 35.33 \\
Gemma-2-9B-Instruct + FUZZY & 24.77 & 32.26 \\
\multicolumn{3}{l}{\textbf{MWOZ dev (FT-mw)}} \\
Gemma-2-9B-Instruct &22.36 & 33.99 \\
Gemma-2-9B-Instruct + FUZZY &25.62& 30.89 \\
\bottomrule
\end{tabular}
\end{table}

The Gemma models immediately surpass SOTA on the SpokenWoZ test set with 38.76\% JGA before post-processing, which yields an another 3\% absolute increase to 42.17\% JGA, achieving the best result among the models presented in this work. However, the FT-mw counterparts of these models do not achieve nearly as stellar performance on the MWOZ dev and human-verbatim test sets, indicating further room for improvement in terms of generalization to previously unseen named entities in the MWOZ test data.

\section{Conclusions}

In this work, we propose an end-to-end dialogue state tracking system based on bridging the representation spaces of a~pretrained speech encoder with an LLM via a~small transformer connector with a~two-step ASR-DST fine-tuning scheme.
We use open source models for both the encoder (WavLM-large) and the LLM (OLMo-1B) to mitigate the risk of test data contamination which would otherwise obstruct meaningful analysis of the scheme.
The best OLMo-1B aligned model achieves 34.66\% joint goal accuracy on the SpokenWoZ test set after post-processing with fuzy matching, significantly outperforming prior work on this dataset.
Finally, experiments conducted with the open-weights Gemma-2-9B-Instruct model yield the best result in this work, achieving 42.17\% JGA on the SWOZ test set.


\section{Acknowledgements}
The work was supported by European Union’s Horizon Europe project No. SEP-210943216 "ELOQUENCE", European Defence Fund project ARCHER, Czech Ministry of Interior project No. VK01020132 "112" and by Czech Ministry of Education, Youth and Sports (MoE) through the OP JAK project "Linguistics, Artificial Intelligence and Language and Speech Technologies: from Research to Applications" (ID:CZ.02.01.01/00/23\_020/0008518). Computing on IT4I supercomputer was supported by MoE through the e-INFRA CZ (ID:90254).

\bibliographystyle{IEEEtran}
\bibliography{authors_wrapper,bibliography}

\end{document}